# Tracer diffusion in colloidal gels

Sujin Babu, Jean Christophe Gimel,* and Taco Nicolai
*Polymères Colloïdes Interfaces, CNRS UMR6120,
Université du Maine, F-72085 Le Mans cedex 9, France*
(Dated: May 7, 2007)

Computer simulations were done of the mean square displacement (MSD) of tracer particles in colloidal gels formed by diffusion or reaction limited aggregation of hard spheres. The diffusion coefficient was found to be determined by the volume fraction accessible to the spherical tracers ($\phi_a$) independent of the gel structure or the tracer size. In all cases, critical slowing down was observed at $\phi_a \approx 0.03$ and was characterized by the same scaling laws reported earlier for tracer diffusion in a Lorentz gas. Strong heterogeneity of the MSD was observed at small $\phi_a$ and was related to the size distribution of pores.

PACS numbers:

Irreversible aggregation of colloidal particles such as proteins [1], clay [2] or oil droplets [3] in solution often leads to the formation of a percolating structure that can resist stress. Recently, the colloidal gel formation has been studied in detail for diffusion limited cluster aggregation of hard spheres using off-lattice computer simulations [4] and [5]. The gels have locally a self-similar structure characterized by a fractal dimension and are homogeneous beyond a characteristic length scale that decreases with increasing volume fraction of the particles ($\phi$).

The transport properties of tracer particles in colloidal gels obviously depend on the volume fraction of the gels that is accessible to the particles ($\phi_a$). The accessible volume, or porosity, depends on the size of the tracers; if the tracers are very small compared to the colloids, $\phi_a$ is close to $1 - \phi$ [6], but it decreases for a given $\phi$ with increasing size of the tracers. Consequently, the long time diffusion coefficient ($D$) of the tracers decreases with increasing $\phi$ or tracer size and goes to zero at a critical value of $\phi_a$. When the accessible volume is small it consists of randomly branched pores that can be of finite size or else percolate through the system.

It has long been known that transport close to the dynamical arrest can be described in terms of diffusion on a percolating network [7]. The geometrical and transport properties of percolation have been investigated extensively on lattices using computer simulations [8] [9] and [10]. The diffusion coefficient of particles was found to go to zero at the percolation threshold following a power law: $D \propto \varepsilon^\mu$, where $\varepsilon = (\phi_a - \phi_a^c)/\phi_a^c$ is the relative distance of the accessible volume fraction to the threshold value ($\phi_a^c$). Close to the threshold the mean square displacement (MSD) of the tracers becomes sub-diffusive meaning that the MSD has a power law dependence on time: $\langle r^2 \rangle \propto t^k$ with $k < 1$ [7] and [11].

In standard lattice simulations the probability to move between neighbouring sites of the network is constant. However, in real systems the pores have a broad continuous range of channel diameters and therefore the local mobility of tracers varies in space. In order to account for this effect, the lattice model was extended to include a power law distribution of probabilities to move between sites [8]. If the exponent of this power law is less than unity, the mobility over large distances is dominated by the lowest probability, which decreases with increasing distance. For this reason $\mu$ and $k$ are reduced to an extent that depends on the exponent. Two different estimates of $\mu$ and $k$ were given for the case of randomly distributed overlapping spherical obstacles, i.e. obstacles forming a so-called Lorentz gas, leading to slightly different values of $\mu$ and $k$: 2.38 and 0.36 [8] or 2.88 and 0.32 [9].

Recently, detailed off-lattice simulations were reported on the tracer diffusion in porous media formed by a Lorentz gas at different densities very close to the percolation threshold. [11]. The aim was to verify the adequacy of the extended lattice model for this system. Anomalous diffusion was observed at the threshold and the exponents $\mu$ and $k$ were found to be consistent with the predictions by Machta et al [9]. The critical value of the accessible volume fraction was $\phi_a^c = 0.0298$ close to values found with other simulations [10] and in experiments on real systems [12].

However, a Lorentz gas of overlapping spherical obstacles is not a realistic model for particle gels. Here we investigate the transport in particle gels formed by irreversible aggregation of hard spheres using computer simulations. Two limiting cases are gels formed by diffusion limited aggregation (DLCA) in which a rigid bond is formed at each collision and reaction limited aggregation (RLCA) in which the bond formation probability goes to zero. Irreversible aggregation leads to the formation of self-similar aggregates with a fractal dimension 1.8 for DLCA and 2.1 for RLCA [13]. When the aggregates have grown to the extent that they fill up the space they connect into a system spanning structure. Such gels can actually be made and the diffusion of tracer particles in such systems can be determined experimentally using e.g. confocal laser scanning microscopy [14] or pulsed field gradient NMR [15]. In order to investigate the ef-fect of spatial correlation caused by aggregation, we also studied tracer self-diffusion in systems of frozen randomly distributed hard spheres (FHS).

---

*Electronic address: Jean-Christophe.Gimel@univ-lemans.fr

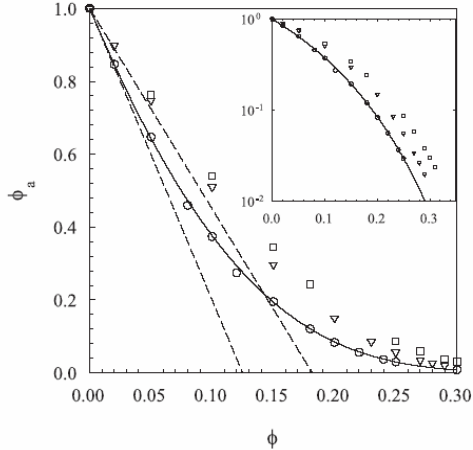

FIG. 1: The accessible volume fraction as a function of $\phi$ for tracer spheres in FHS ($\bigcirc$), and DLCA ($\triangle$) and RLCA ($\square$) gels of spheres with the same diameter as the tracers. The inset shows the same data on a logarithmic scale for $\phi_a$. The solid line represents values calculated using the Carnahan-Starling equation. The dashed lines represent the limiting low concentration behaviour, see text.

Gels were simulated by irreversible cluster-cluster aggregation starting from a random distribution of hard spheres with unit diameter until all particles are connected, see [4] for details. The diffusion of tracers was simulated by small displacements in a random direction. If the displacement led to overlap the movement was either refused or truncated at contact. The step size was chosen sufficiently small so that reducing it further had no significant effect on the results. The time unit was chosen as the time needed for a tracer to diffuse over its diameter at infinite dilution, which is about $0.4s$ for a particle with a diameter of $1\mu m$ in water at room temperature. Simulations were done in a box with length 50 using periodic boundary conditions. We checked for finite size effects by varying the box size, and all results shown here are not influenced by finite size effects. We averaged over several hundred paths of randomly inserted tracers and 10 independent system configurations. $\phi_a$ was calculated as the probability that a tracer could be randomly inserted without overlap.

We present first results for the case that the tracer particles have the same diameter ($d$) as the obstacle particles, after which we discuss the effect of varying the tracer size. In Fig. 1 the dependence of $\phi_a$ is plotted as a function of $\phi$ for randomly distributed hard spheres and gels formed by DLCA and RLCA. In the latter case, $\phi_a$ increased with decreasing bond formation probability, but the variation became negligible below $10^{-4}$, which we have taken as the RLCA limit. For a given concentration, $\phi_a$ is larger for RLCA gels than for DLCA gels which in turn is larger than for the hard sphere system. Gels have a larger $\phi_a$, because the particles are connected and therefore have a larger fraction of overlapping excluded volume. $\phi_a$ is larger for RLCA than DLCA, because RLCA clusters are denser. For non-interacting hard spheres $\phi_a$ is directly related to the chemical potential ($\mu_{cs}$): $\phi_a = \phi \exp(-\mu_{cs})$ [16], and can be calculated using the so-called Carnahan-Starling equation [17] for $\mu_{cs}$, see solid line in Fig 1. For randomly distributed hard spheres the initial dependence of $\phi_a$ on $\phi$ is given by: $\phi_a = 1 - \phi(1+b)^3$, where $b$ is the size ratio of the tracers over the obstacles [6]. For gels the initial dependence can be estimated by assuming that the gels consist of strands of touching spheres: $\phi_a = 1 - \phi(1 + 3b + 1.5b^2)$. The dashed lines in Fig 1 show that these estimates are only valid for small $\phi_a$.

Images of the accessible volume in DLCA gels at different $\phi$ are shown in Fig. 2. At low volume fractions (Fig. 2a), almost all pores percolate through the system (*yellow*), but with increasing $\phi$ (Figs 2b and 2e) the fraction of finite pores increases until above a critical value ($\phi^c$) (Fig. 2d) there is no longer a percolating pore. For clarity, we have shown the percolating pore separately in Figs 2c and 2f.

The MSD averaged over all tracers is shown in Fig. 3 for DLCA gels at different $\phi$. The results are similar to those obtained by Höfling et al [11] for the Lorentz gas. Initially, the tracers diffuse freely until they hit the obstacles. Then the displacement of tracers is anomalous until $\langle r^2 \rangle$ exceeds a characteristic value ($\xi^2$) after which it becomes again diffusional with a reduced diffusion coefficient. $\xi$ represents the correlation length of the percolating pores and diverges at the threshold. The tracers in finite size pores are trapped and do not contribute to $\langle r^2 \rangle$ at long times. For $\phi > \phi_c$, all the tracers are trapped and $\langle r^2 \rangle$ stagnates at twice the averaged squared radius of gyration of the pores ($\langle R_g^2 \rangle$). This follows from the fact that the average distance a tracer has moved in a finite size pore at $t \to \infty$ is the same as the average distance between two randomly placed tracers.

In Fig. 4, the dependence on $\phi$ of the long time diffusion coefficient relative to the free diffusion coefficient ($D$) is compared for DLCA and RLCA gels and frozen random hard spheres. For a given volume fraction, $D$ is close for the two gels, but is smaller for FHS. We note that for systems of freely moving obstacles $D$ decreases much more slowly with increasing $\phi$ [18]. $D$ goes to zero at $\phi_c$ equal to $0.248 \pm 0.003$, $0.279 \pm 0.001$ and $0.295 \pm 0.005$, for FHS, DLCA and RLCA, respectively.

The same results are plotted as a function of $\phi_a$ in Fig. 5. For comparison we have included in Fig. 5 results obtained for the Lorentz gas from [11]. Here, and in [11], only the diffusion of tracers in the pores is considered.

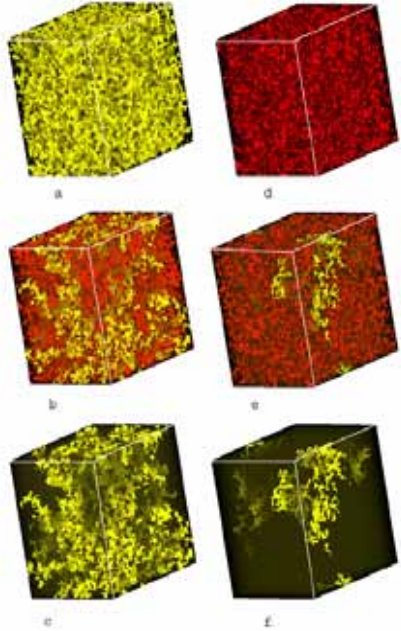

FIG. 2: Images of the accessible volume for DLCA gels at different values of $\phi_a$. Percolating and isolated pores are indicated in yellow and red, respectively. For clarity figures 2c and 2f show just the percolating pore of systems shown in figures 2b and 2e, respectively.

Other results for the Lorentz gas were obtained in the context of the conductivity that is proportional to point tracer diffusion [6] and [19]. In these simulations the average was taken over all tracers including the immobile ones placed in the obstacles so that $D$ is reduced by a factor $\phi_a$. After correction, these results are close to the results shown in Fig. 5. Experimental results derived from conductivity measurements on fused spherical glass beads in water [12] are close to the simulation results. For $\phi_a$ close to unity $D$ should decrease as $\sqrt{\phi_a}$ [20]. The full dependence can be described approximately by the following empirical equation which has the predicted limiting behavior for $\phi_a \to 1$ and $\phi_a \to \phi_a^c$:

$$D = \sqrt{\phi_a} \left[ \frac{\phi_a - \phi_a^c}{\phi_a(1 - \phi_a^c)} \right]^\mu \quad (1)$$

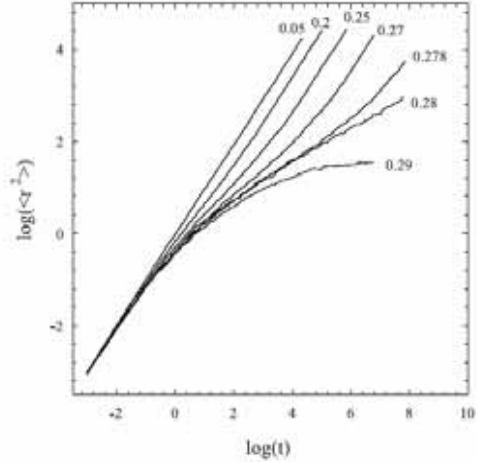

FIG. 3: MSD of tracer spheres in DLCA gels of spheres with the same diameter as the tracers at different $\phi$, indicated in the figure.

with $\phi_a^c = 0.03$ and $\mu = 2.8$ see solid lines in Fig. 5.

It appears that $\phi_a$ is the parameter that determines the diffusion coefficient for all these different systems; the variation of $\phi_a$ for a given $D$ is less than 20%. In each case $D$ goes to zero at a critical value ($\phi_a^c$) close to 0.03, but the actual value of $\phi_a^c$ is not universal as noted earlier by Rintoul [10]. The inset of Fig. 5 shows the data plotted as a function of the distance to the percolation threshold $\varepsilon$. For the Lorentz gas we have used the precise value of $\phi_a^c$ calculated by Rintoul: $\phi_a^c = 0.0301 \pm 0.0003$. For the other systems we do not have the same precision, but we found that the data superimpose close to $\phi_a^c$ if we choose $\phi_a^c = 0.029$ for FHS, $\phi_a^c = 0.02655$ for DLCA gels and $\phi_a^c = 0.033$ for RLCA gels.

The dependence of $D$ close to the percolation threshold is compatible with a power law: $D \propto \varepsilon^\mu$. However, there is considerable uncertainty in the value of the exponent due to the strong correlation between $\phi_a^c$ and $\mu$. For instance, Höfling et al. [11] fixed $\mu$ at 2.88 predicted by Machta et al. [9] and found in this way $\phi_a^c = 0.0298$. Fixing $\phi_a^c$ at 0.0301 we find $\mu = 2.5$. In fact predictions for $\mu$ from [8] and [9] are both compatible with the data, see Fig. 5, and we are not in the position to decide which, if any, is correct. Nevertheless, $\mu$ is clearly larger than the value 1.88 obtained from lattice simulations of diffusion in percolating systems [7].

The power law exponent, $k$, of the anomalous MSD at the threshold is related to $\mu$ as [7]: $k = 2(\nu - \beta/2)/(2\nu + \mu - \beta)$, where $\beta$ characterizes the dependence of the volume fraction of the percolating pores ($\phi_a^p$) close to the threshold: $\phi_a^p \propto \varepsilon^\beta$ and $\nu$ characterizes the divergence of the correlation length: $\xi \propto \varepsilon^{-\nu}$. If one only considers the displacement of tracers in the percolating pore the exponent is larger: $k' = 2\nu/(2\nu + \mu - \beta)$. Fig. 6 compares the average MSD at the percolation threshold of tracers placed anywhere in the accessible volume with that of tracers placed in the percolating pore. Utilizing the values for $\nu = 0.88$ and $\beta = 0.41$ obtained from lattice simulations [21] gives $k = 0.36$ and $k' = 0.47$ for $\mu = 2.38$ and $k = 0.32$ and $k' = 0.42$ for $\mu = 2.88$. The latter appear to describe the data better, but unfortunately, accurate determination the limiting power law behaviour is largely beyond the current computer capacities, given the fact that the limiting power law behaviour of the cluster size distribution is not yet observed even for lattice simulations with a box size of 1023 [22].

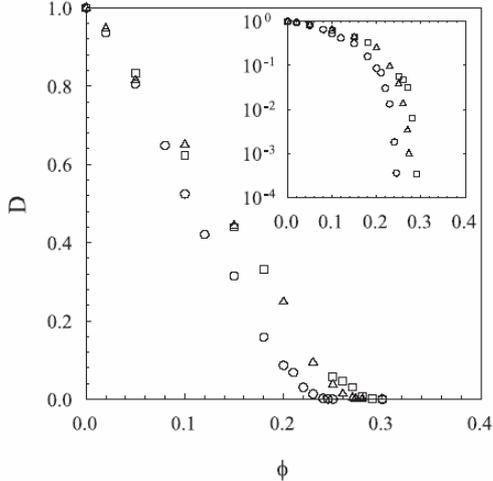

FIG. 4: Relative diffusion coefficient of tracer spheres as a function of $\phi$ in FHS ($\bigcirc$), and DLCA ($\triangle$) and RLCA ($\square$) gels of spheres with the same diameter as the tracers. The inset shows the same data on a logarithmic scale for $D$.

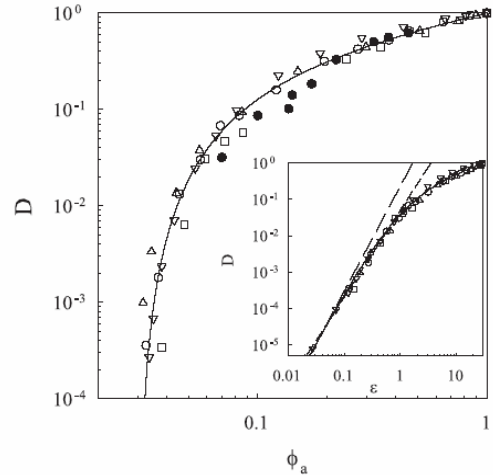

FIG. 5: Relative diffusion coefficient of tracer spheres as a function of $\phi_a$ in FHS ($\bigcirc$), and DLCA ($\triangle$) and RLCA ($\square$) gels of spheres with the same diameter as the tracers. The results for the Lorentz gas from [11] ($\triangledown$) and experimental from [12] ($\bullet$) are also shown for comparison. The solid line represents Eq. 1 where we have used $\phi_a^c = 0.03$ and $\mu = 2.8$. The inset shows the same data plotted as a function of $\varepsilon = (\phi_a - \phi_a^c)/\phi_a^c$ on a logarithmic scale. The straight lines in the inset represent the predicted power law dependences from [8] (short dashed) and [9] (long dashed).

The displacement of tracers is highly heterogeneous close to the percolation threshold and can be characterized in terms of the probability distribution that tracers have moved a distance $r^2$ at time $t$ ($P(r^2)$). For $\phi_a > \phi_a^c$, we need to distinguish between the fraction ($\phi_a^p$) of tracers in the percolating pore and the fraction ($1 - \phi_a^p$) in finite size pores. $P(r^2)$ of tracers in the percolating pores is Gaussian if the MSD is much larger than $\xi^2$ and $\langle r^2 \rangle$ increases linearly with time. On the other hand, $P(r^2, t)$ of tracers trapped in a finite size pore becomes independent of $t$ and $\langle r^2 \rangle$ stagnates at $2\langle R_g^2 \rangle$. Therefore $P(r^2)$ splits up into two peaks: $P_1$ representing tracers in the percolating pore that displaces linearly with $t$ and $P_2$ representing tracers in finite size pores that stagnates, with amplitudes $\phi_a^p$ and $(1 - \phi_a^p)$, respectively. The split-up is illustrated in Fig.7a where $P(r^2)$ is shown at three times for a DLCA gel just above the percolation threshold ($\phi_a = 0.0276$). At the shortest time the MSD is still much smaller than $\xi^2$, and one observes a single distribution. The split-up starts when $\langle r^2 \rangle \approx \xi^2$ and is clearly visible at the longest time when $\langle r^2 \rangle$ is much larger than $\xi^2$. Fig. 7b shows the distributions at $t = 10^6$ for three values of $\phi_a$. At $\phi_a = 0.0265$, i.e. smaller than $\phi_a^c$, only one peak is observed representing the pore size distribution, which is broad close to the threshold. At $\phi_a = 0.056$ almost all accessible volume percolates and a narrow peak is seen that shifts to larger $r^2$ with time. For $\phi_a = 0.032$, i.e. just above $\phi_a^c$, both peaks representing freely diffusing and trapped tracers are observed.

The same features were found when $\phi_a$ was varied by varying the tracer size at constant obstacle volume fraction. As mentioned above, $\phi_a$ decreases with increasing tracer size starting from $\phi_a = 1 - \phi$ for point tracers see inset of Fig. 8. Examples of the dependence of $D$ on $\phi_a$ are given in Fig. 8 for different tracer diameters between 0.1 and 1 for gels and FHS at fixed volume fractions. For the Lorentz gas there is a strict equivalence between the diffusion of point tracers and finite size tracers at the same $\phi_a$ [6]. Similar equivalence exists between finite size tracers in FHS and point tracers in randomly distributed semi-penetrable spheres. It was found, for a limited range of $\phi_a$, that $D$ was the same for a given $\phi_a$ for tracers with different sizes in FHS [6]. Fig. 8 shows that the effect of tracer size on $D$ is essentially determined by $\phi_a$ also for gels, but the relationship is not exact.

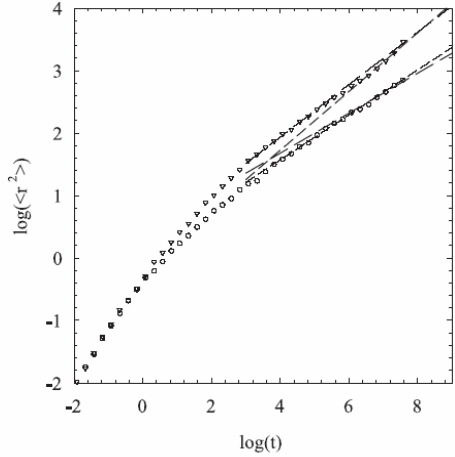

FIG. 6: Comparison of the MSD of tracers in a DLCA gel at the percolation threshold placed anywhere in the accessible volume (○) or just in the percolating pore (▽). The straight lines represent the predicted power law dependences in the system and in the percolating pore from [8] (short dashed) and [9] (long dashed).

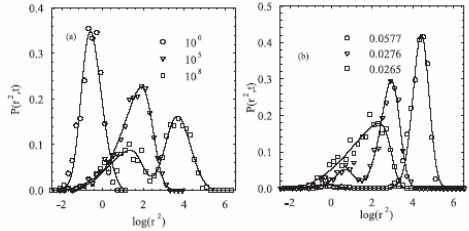

FIG. 7: Distribution of the MSD of tracers in a DLCA gel for a fixed $\phi_a = 0.0276$ close to the percolation threshold at different times (a) and for a fixed $t = 10^6$ at different $\phi_a$ (b), as indicated in the figure.

In conclusion, tracer diffusion in colloidal gels is slower than in systems of freely diffusing hard spheres, but faster than in frozen randomly distributed spheres. The latter effect is due to the increase of the accessible volume when the particles gel. $D$ is mainly, but not fully, determined by $\phi_a$ independent of the gel structure and the tracer size. $\phi_a$ can thus be deduced from a measurement of $D$. The tracer diffusion becomes zero at a critical value of accessible volume $\phi_a \approx 0.03$ that is almost the same for gels and frozen hard spheres. The dependence of $D$ on $\phi_a$ close to $\phi_a^c$ can be described in terms of a power law dependence on the distance to $\phi_a^c$. The MSD at $\phi_a^c$ is anomalous and increases as a power law with time with an exponent less than unity. The exponents of the two power law relationships are related and are consistent with the percolation model.

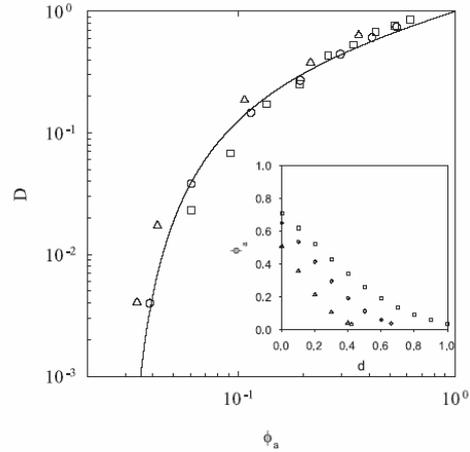

FIG. 8: Relative diffusion coefficient of tracer spheres with different diameters between 0.1 and 1 as a function of $\phi_a$ in FHS at $\phi = 0.35$ (circle), a DLCA gel at $\phi = 0.49$ (triangle) and a RLCA gel at $\phi = 0.29$ (square) formed by spheres of unit diameter. The solid line is the same as in Fig. 5. The inset shows the dependence of $\phi_a$ on the tracer diameter.

### Acknowledgments

This work has been supported in part by a grant from the Marie Curie Program of the European Union numbered MRTN-CT-2003-504712.